\documentclass[showpacs,nopreprintnumbers,10pt,a4paper,notitlepage,nofootinbib,aps,prd,amsfonts,amssymb,amsmath]{revtex4-2}

\usepackage{hyperref}
\usepackage{graphicx}

\begin{document}

\title{Non-relativistic stellar structure in the Fierz--Pauli theory and generic linear massive gravity}

\author{Tomoya Tachinami}\email[]{tachinami(a)tap.st.hirosaki-u.ac.jp}
\affiliation{
Graduate School of Science and Technology, Hirosaki University,
Hirosaki, Aomori 036-8561, Japan
}
\author{Yuuiti Sendouda}\email[]{sendouda(a)hirosaki-u.ac.jp}
\affiliation{
Graduate School of Science and Technology, Hirosaki University,
Hirosaki, Aomori 036-8561, Japan
}

\date{\today}

\begin{abstract}
We study the structure of static spherical stars composed of non-relativistic matter in linear massive gravity with or without the Fierz--Pauli (FP) tuning.
Adopting a polytropic equation of state, we construct master differential equations for the stellar profile function, which is fourth order in the FP theory or sixth order in generic non-FP theories, where the difference in the differential order reflects the presence of a ghost spin-$ 0 $ graviton in the latter.
In both cases, even when the spin-$ 0 $ ghost is present, we find exact solutions with finite radius for the polytropic indices $ n = 0 $ and $ 1 $.
Analyzing the dependences of the stellar radius, mass, and Yukawa charge on the graviton masses, we observe that a discontinuous behavior arises in the massless limit of the FP theory similarly to the van Dam--Veltman--Zakharov discontinuity, while it is absent in non-FP theories.
We discuss rough observational constraints on the graviton masses.
\end{abstract}

\maketitle

\section{Introduction}

General relativity (GR), a gauge theory of a massless spin-$ 2 $ graviton, has long established the status of the standard, most successful model of gravity to date.
There are a great number of successes from the observation of the gravitational bending of light in 1919 to the first detection of gravitational waves in 2015.
Nonetheless, there remain theoretical and observational challenges that forces us not to reach a consensus that it is a perfect model.
One of the most challenging problems confronting GR concerns the origin of the late-time acceleration of the Universe, which, in the framework of GR, is customarily ascribed to some dark energy, an unknown energy component fulfilling the Universe with negative pressure.
Although there have been plenty of hypothetical candidates, such as a cosmological constant, quintessence, and so on, none of them seems decisive so far.

Instead, in this paper, we take a strategy to explore another class of possibilities which involves modifications to GR.
In such a cosmological context, massive gravity (MG), the idea to give a mass to the graviton, has gathered much interests recently.
Massiveness of the graviton, if any, would introduce a length scale to gravitational interaction, weakening the attraction of gravity at large distances so that the Universe gets able to accelerate.
At present, however, the search for a plausible nonlinear theory of MG is still underway.
The recent remarkable developments are reviewed in \cite{deRham:2014zqa}.

Typically, models of nonlinear MG, when perturbed around the flat spacetime, settle down to a class of linear theories specified by an action being second order in the metric perturbation $ h_{\mu\nu} \equiv g_{\mu\nu} - \eta_{\mu\nu} $\,,
\begin{equation}
S_\mathrm g
= \frac{1}{32 \pi\,G}\,\int\!\mathrm d^4x\,
  \left[
   -\frac{1}{2}\,\partial_\rho h_{\mu\nu}\,\partial^\rho h^{\mu\nu}
   + \partial_\rho h_{\mu\nu}\,\partial^\mu h^{\nu\rho}
   - \partial_\mu h^{\mu\nu}\,\partial_\nu h
   + \frac{1}{2}\,\partial_\mu h\,\partial^\mu h
   - \frac{m^2}{2}\,\left(h_{\mu\nu}\,h^{\mu\nu}-(1-\epsilon)\,h^2\right)
  \right]\,,
\label{eq:action_g}
\end{equation}
where $ G $ is the (bare) Newton constant and $ m $ is a mass parameter,\footnote{We work with the natural units such that $ c = \hbar = 1 $.} which is missing in GR.
In order to enjoy impeccable successes of GR in weak regimes, the above linear theory necessarily involves a spin-$ 2 $ graviton, with an essential difference being the non-zero mass which breaks the gauge symmetry of GR.
Here, the parameter $ \epsilon $ is what we shall call the ``non-Fierz--Pauli'' parameter.
It is known that a non-zero value of $ \epsilon $ signals the emergence of another massive graviton of spin-$ 0 $, which is a ghost mode with negative kinetic energy \cite{}.
The theory without the ghost, $ \epsilon = 0 $, is actually the one proposed by Fierz and Pauli (FP) in 1939 \cite{Fierz:1939ix}, which is the most traditional and simplest model of linear MG.

A crucial observation associated with the FP theory with a matter field was made by van Dam and Veltman and by Zakharov in 1970 \cite{vanDam:1970vg,Zakharov:1970cc}.
They added a minimally coupled matter source
\begin{equation}
S_\mathrm{int}
= \frac{1}{2}\,\int\!\mathrm d^4x\,h_{\mu\nu}\,T^{\mu\nu}
\label{eq:action_m}
\end{equation}
to the gravitational action \eqref{eq:action_g} with $ \epsilon = 0 $.
Their discovery was that the prediction for the bending angle of light due to a massive body in the massless limit of the FP theory differs from GR by a factor of $ 3/4 $, a phenomenon called the van Dam--Veltman--Zakharov (vDVZ) discontinuity.
Rather than rejecting the FP theory, however, the discovery has stimulated inspections of ``screening'' mechanisms in MG that can cancel the discrepancy and make predictions on the solar-system scale consistent with GR.
One of the most notable proposals was made by Vainshtein \cite{Vainshtein:1972sx}, who considered a non-linear kinetic term and found recovery of consistency with GR within a certain radius.
See \cite{Hinterbichler:2011tt} for historical overviews and recent developments.

The purpose of this study is to perform another test of linear MG to examine what roles are played by the mass $ m $ as well as the non-FP parameter $ \epsilon $ in astrophysical bodies.
Specifically, we will develop differential equations determining the density profile of a non-relativistic spherically symmetric star in hydrostatic equilibrium in linear MG theories with or without the FP ``tuning'' $ \epsilon = 0 $.
We will adopt the polytropic relation as the equation of state (EOS) and try to obtain some analytic solutions for the polytropic index $ n = 0 $ and $ 1 $ in all MG theories, which can be compared with the results in GR.
In doing so, we will take advantage of our recent work \cite{Tachinami:2021jnf} in identifying independent dynamical degrees of freedom (DOFs) in the non-FP theories.
We also note that the present scheme has a great similarity to the one employed in the study of stellar structures in higher-curvature gravity in Ref.~\cite{Tonosaki:2023trc}.

The fundamental question is what range of the graviton mass $ m $ is allowed from the perspective of stellar structures.
A severe upper bound has been recently placed on the deviation from GR, i.e., finiteness of the graviton mass, from the multi-messenger observation of a gravitational-wave event GW170817 \cite{LIGOScientific:2017zic}.
However, the experience of the vDVZ discontinuity cautions that something inconsistent with GR may occur in the massless limit of the FP-tuned theory with $ \epsilon = 0 $.
In this paper, we shall be greatly cautious about this limit and explore any different consequences arising when the FP tuning is violated at the risk of having the spin-$ 0 $ ghost.
Our current standpoint regarding the ghost is that it is not necessarily forbidden in a classical linear theory and the non-FP MG may be accepted as a phenomenological effective theory.

The organization of this paper is as follows.
In Sec.~\ref{sec:lane}, we derive a modified Lane--Emden (LE) equation, which is a master equation for non-relativistic stellar structures in MG.
We will also treat limiting cases recovering the FP theory and GR.
In Sec.~\ref{sec:esol}, we will solve the modified LE equation with the polytropic indices $ n = 0 $ and $ 1 $.
We try to find its exact solutions and study the dependence of their stellar radius, mass, and Yukawa charge on the graviton masses.
Semi-observational constraints on the graviton masses are also discussed.
Finally we conclude in Sec.~\ref{sec:concl}.

Throughout the paper, we will work with the natural units with $ c = \hbar = 1 $.
Greek indices of tensors such as $ \mu,\nu,\cdots $ are of space-time while Latin ones such as $ i,j,\cdots $ are spatial.
The Minkowski metric is $ \eta_{\mu\nu} $.
The symbol $ \partial_\mu $ denotes partial differentiation $ \frac{\partial}{\partial x^\mu} $\,.
$ \square \equiv \eta^{\mu\nu}\,\partial_\mu \partial_\nu $ is the d'Alembertian and $ \triangle \equiv \delta^{ij}\,\partial_i \partial_j $ the Laplacian.

\section{\label{sec:lane}Hydrostatic equilibrium and the modified Lane--Emden equation}

In this section, we derive an equation determining the structure of a newtonian star in hydrostatic equilibrium in generic linear MG that can be viewed as a modified version of the Lane--Emden (LE) equation in GR.

\subsection{Scalar-type equations of motion in MG}

We begin with summarizing the equations of motion for the scalar modes in MG, which will be relevant to static spherical configurations.
Scalar-type metric perturbations are characterized by four variables as
\begin{equation}
h_{00}^{(\mathrm S)}
= -2 A\,,
\quad
h_{0i}^{(\mathrm S)}
= -\partial_i B\,,
\quad
h_{ij}^{(\mathrm S)}
= 2 \delta_{ij}\,C + 2 \partial_i \partial_j E\,.
\end{equation}
It is well known that one can choose two linearly independent combinations of the above four variables that are invariant under a small gauge transformation, see Appendix~\ref{sec:gauge} for a summary of linear gauge transformations and invariant variables.
In particular,
\begin{equation}
\Psi
\equiv
  A - \dot B - \ddot E\,,
\quad
\Phi
\equiv
  C
\label{eq:giv_scalar}
\end{equation}
correspond to the gravitational potentials in the Newtonian gauge.
When the action \eqref{eq:action_g} does not possess the mass term, the scalar sector can be solely written in terms of the gauge-invariant variables because the massless theory inherits the linearized version of the gauge symmetry of GR.
In the massive case, on the other hand, one cannot take such advantage because the mass term breaks the symmetry.
Indeed, the action \eqref{eq:action_g} is written in terms of the metric perturbations, $ A $, $ B $, $ C $, and $ E $, as
\begin{equation}
\begin{aligned}
S_{\mathrm g}^{(\mathrm S)}
&
= \frac{1}{16 \pi\,G}\int\!\mathrm d^4x\,\biggl[
   -6 C\,\square C - 4C\,\triangle (A-\dot B-\ddot E-C) \\
& \qquad
   - \frac{m^2}{2}\,\left(
      2 \epsilon\,A^2
      + B\,\triangle B
      + 6 (3\epsilon-2)\,C^2
      + 2 \epsilon\,E\,\triangle^2 E
      + 4 (3\epsilon-2)\,C\,\triangle E
      + 4 (\epsilon-1)\,A\,\triangle E
      + 12 (\epsilon-1)\,A\,C
     \right)
  \biggr]\,,
\end{aligned}
\end{equation}
up to surface terms.
In a similar fashion, we define the scalar components of the perturbative energy-momentum tensor $ T_{\mu\nu} $ as
\begin{equation}
T_{00}^{(\mathrm S)}
= \rho\,,
\quad
T_{0i}^{(\mathrm S)}
= -\partial_i v\,,
\quad
T_{ij}^{(\mathrm S)}
= P\,\delta_{ij}
  + \left(\partial_i \partial_j - \frac{1}{3}\,\delta_{ij}\,\triangle\right)\,\sigma\,,
\end{equation}
so the scalar sector of the interaction action \eqref{eq:action_m} is
\begin{equation}
S_\mathrm{int}^{(\mathrm S)}
= \int\!\mathrm d^4x\,\left[
   -A\,\rho
   + B\,\triangle v
   + 3 C\,P
   + E\,\triangle\,\left(P + \frac{2}{3} \triangle\,\sigma\right)
  \right]\,,
\end{equation}
where surface terms have been discarded.
We assume the conservation law $ \partial_\mu T^{(\mathrm S)}{}^{\mu\nu} = 0 $ holds, which settles down in the decomposed form:
\begin{equation}
\dot\rho + \triangle v = 0\,,
\quad
\dot v + P + \frac{2}{3}\,\triangle\sigma = 0\,.
\end{equation}
Varying the action $ S^{(\mathrm S)} = S_\mathrm{g}^{(\mathrm S)} + S_\mathrm{int}^{(\mathrm S)} $ with respect to $ A\,,B\,,C\,, E $ and using the conservation law, we obtain the four equations of motion (EOMs)
\begin{equation}
\begin{aligned}
2 \triangle C + m^2\,[\epsilon\,A+(\epsilon-1)\,\triangle E+3(\epsilon-1)\,C]
&
= -8 \pi\,G\,\rho\,,\\
\triangle (4 \dot C+m^2\,B)
&
= -16 \pi\,G\,\dot\rho\,,\\
6 \square C + 2 \triangle (A-\dot B-\ddot E-2C)
+ m^2\,[3(3\epsilon-2)\,C+(3\epsilon-2)\,\triangle E+3(\epsilon-1)\,A]
&
= 24 \pi\,G\,P\,,\\
\triangle \{-2\ddot C+m^2\,[\epsilon\,\triangle E+(3\epsilon-2)\,C+(\epsilon-1)\,A]\}
&
= 8 \pi\,G\,\ddot\rho\,.
\end{aligned}
\label{eq:eom}
\end{equation}
The derivation of an equivalent set of equations can be found in Ref.~\cite{Jaccard:2012ut}.

The authors rigorously proved in Ref.~\cite{Tachinami:2021jnf} that, in the vacuum case, the scalar-type dynamical content in this theory consists of the helicity-$ 0 $ component of the spin-$ 2 $ graviton and the spin-$ 0 $ graviton, which may be defined as
\begin{equation}
\phi_2
\equiv
  \frac{1}{2}\,(A - \dot B - \ddot E - C)\,,
\quad
\phi_0
\equiv
  -A + \frac{\dot B}{2} - 2 C\,,
\label{eq:phi_def}
\end{equation}
respectively.
In fact, even in the presence of the matter sources, the EOMs \eqref{eq:eom} are neatly recast into the following two sourced Klein--Gordon-type equations for $ \phi_2 $ and $ \phi_0 $\,,
\begin{equation}
(\square - m_2^2)\,\phi_2
= 4 \pi\,G\,(\rho + \dot v - \square\sigma)\,,
\quad
(\square - m_0^2)\,\phi_0
= 4 \pi\,G\,(\rho - 3 P)\,,
\end{equation}
where their masses are defined as
\begin{equation}
m_2^2
\equiv
  m^2\,,
\quad
m_0^2
\equiv
  \frac{3 - 4\epsilon}{2\epsilon}\,m^2\,,
\end{equation}
respectively.
By solving $ \phi_2 $ and $ \phi_0 $\,, we can reconstruct the original metric variables, see Appendix~\ref{app:metric} for the formula.

In order to prevent $ \phi_0 $ from being tachyonic, the non-FP parameter $ \epsilon $ is required to be within the range $ 0 < \epsilon \leq 3/4 $ so that $ 0 \leq m_0^2 < \infty $.
We will shortly confirm that both $ \phi_2 $ and $ \phi_0 $ produce a Yukawa-type potential in the static case and, when non-tachyonic, $ \phi_2 $ is attractive in the remote distances whereas $ \phi_0 $ is repulsive.
On the observational ground, the attractive force mediated by the spin-$ 2 $ graviton must dominate, so we further restrict the range of $ \epsilon $ within $ 0 < \epsilon \leq 1/2 $ so that $ m_2^2 \leq m_0^2 $\,.
On the other hand, $ m_2^2 $ must be so tiny that the speed of gravitational waves is sufficiently close to the speed of light in light of the observation of GW170817 \cite{LIGOScientific:2017zic}.
Notice that, as long as $ \epsilon \neq 0 $, vanishing of the spin-$ 2 $ mass implies simultaneous vanishing of the spin-$ 0 $ mass.

\subsection{Massive gravitational potentials: Absence of the vDVZ discontinuity}

From now on, we specialize to static configurations with non-relativistic matter satisfying $ \rho \gg |P| $.
Then, our variables are related to the metric perturbations as
\begin{equation}
\phi_2
= \frac{A - C}{2}\,,
\quad
\phi_0
= -A - 2 C
\end{equation}
and they obey the Helmholtz-type equations
\begin{equation}
(\triangle-m_2^2)\,\phi_2
= 4 \pi\,G\,\rho\,,
\quad
(\triangle-m_0^2)\,\phi_0
= 4 \pi\,G\,\rho\,.
\label{eq:Helmholtz}
\end{equation}
The gauge-invariant variables \eqref{eq:giv_scalar} can now be written in terms of $ \phi_2 $ and $ \phi_0 $ as
\begin{equation}
\Psi
= A
= \frac{4}{3}\,\phi_2 - \frac{1}{3}\,\phi_0\,,
\quad
\Phi
= C
= -\frac{2}{3}\,\phi_2 - \frac{1}{3}\,\phi_0\,.
\label{eq:potential}
\end{equation}
For notational convenience, we introduce
\begin{equation}
\alpha_2
= \frac{4}{3}\,,
\quad
\alpha_0
= -\frac{1}{3}\,,
\quad
\beta_2
= -\frac{2}{3}\,,
\quad
\beta_0
= -\frac{1}{3}
\end{equation}
and write the gauge-invariant potentials as
\begin{equation}
\Psi
= \sum_{s=2,0} \alpha_s\,\phi_s\,,
\quad
\Phi
= \sum_{s=2,0} \beta_s\,\phi_s\,.
\end{equation}
It will be useful to notice that $ \sum_s \alpha_s = \sum_s \beta_s = 1 $.

Hereafter we presume that the system is spherically symmetric.
Since asymptotic flatness requires $ \lim_{|\vec x| \to \infty} \Psi = \lim_{|\vec x| \to \infty} \Phi = 0 $\,, $ \phi_s $ must satisfy $ \lim_{|\vec x| \to \infty} \phi_s = 0 $\,.
The formal spherically symmetric solution to the Helmholtz equation \eqref{eq:Helmholtz} fulfilling the asymptotic boundary condition is given by
\begin{equation}
\phi_s
= -\frac{G}{r}\,\left(\sigma_s(r)\,\cosh m_s\,r + (I_s-\chi_s(r))\,\sinh m_s\,r\right)
\label{eq:phi}
\end{equation}
with
\begin{equation}
\sigma_s(r)
\equiv
  \frac{4\pi}{m_s}\,\int_0^r\mathrm dr'\,r'\,\sinh{m_sr'}\,\rho(r')\,,
\quad
\chi_s(r)
\equiv
  \frac{4\pi}{m_s}\,\int_0^r\mathrm dr'\,r'\,\cosh{m_sr'}\,\rho(r')\,,
\end{equation}
where the integration constant $ I_s $ must be so tuned to kill the exponentially growing mode.
When the matter distribution is confined within a finite stellar radius $ R $, these functions have a constant value outside the star:
$ \sigma_s(r\geq R)=\sigma_s(r=R) \equiv \Sigma_s $\,, $  \chi_s(r\geq R)=\chi_s(r=R) \equiv X_s $\,. 
Then the integration constant is determined as
\begin{equation}
I_s
\equiv
  X_s - \Sigma_s
= \frac{4\pi}{m_s}\,\int_0^R\!\mathrm dr\,r\,\rho(r)\,\mathrm e^{-m_sr}\,.
\label{eq:I}
\end{equation}
Outside the star, $ \phi_s $ reduces to a Yukawa-type potential
\begin{equation}
\phi_s(r \geq R)
= -\frac{G\,\Sigma_s\,\mathrm e^{-m_sr}}{r}\,,
\end{equation}
because $ \sigma_s(r\geq R) = \Sigma_s $ and $ I_s - \chi_s(r\geq R) = -\Sigma_s $\,.
We see that $ \Sigma_s $ plays a role of a Yukawa charge.

Let us gain some insights into the massless limit.
The limiting value of the enclosed charge $ \sigma_s (r) $ is nothing but the enclosed mass,
\begin{equation}
\lim_{m_s \to 0} \sigma_s(r)
= 4\pi\,\int_0^r\mathrm dr'\,r'^2\,\rho(r')
\equiv
  m(r)\,,
\end{equation}
which proves that the massive potential \eqref{eq:phi} reproduces the Newtonian potential in this limit:
\begin{equation}
\lim_{m_s \to 0} \phi_s(r)
= \int_\infty^r\!\frac{G\,m(r')}{r'^2}\,\mathrm dr'\,.
\end{equation}
A remarkable consequence is that the massless limit of the gauge-invariant potentials external to an object with mass $ M \equiv m(R) $ recovers the Newtonian potential,
\begin{equation}
\lim_{m\to 0} \Psi
= \lim_{m\to 0} \Phi
= -\frac{GM}{r}\,,
\end{equation}
thereby proving the absence of the vDVZ discontinuity in generic non-FP MG.

\subsection{Master equation from hydrostatic equilibrium condition}

Given the gauge-invariant potential $ \Psi $, the hydrostatic equilibrium condition in the newtonian gauge reads
\begin{equation}
\frac{1}{\rho}\,\frac{\mathrm dP}{\mathrm dr}
= -\frac{\mathrm d\Psi}{\mathrm dr}\,.
\label{eq:hydro}
\end{equation}
In order to obtain a differential equation for $ \rho $, one supplies an EOS $ P = P(\rho) $ and operates $ \frac{1}{r^2}\,\frac{\mathrm d}{\mathrm dr} r^2 $ on the both sides to find
\begin{equation}
\frac{1}{r^2}\,
\frac{\mathrm d}{\mathrm dr} \left(\frac{r^2}{\rho}\,\frac{\mathrm dP}{\mathrm dr}\right)
= -\triangle \Psi\,.
\end{equation}
For now, we assume both $ m_2^2 $ and $ m_0^2 $ are bounded.
Then we can safely use the Helmholtz equations \eqref{eq:Helmholtz} for $ \phi_s $'s to obtain
\begin{equation}
\frac{1}{r^2}\,
\frac{\mathrm d}{\mathrm dr} \left(\frac{r^2}{\rho}\,\frac{\mathrm dP}{\mathrm dr}\right)
= -\sum_s \alpha_s\,(m_s^2\,\phi_s + 4 \pi G \rho)\,.
\label{eq:hydro_psi}
\end{equation}

Hereafter we adopt the polytropic relation $ P = K\,\rho^{1+\frac{1}{n}} $ as the EOS for the stellar matter, where $ K $ is a constant and $ n $ is another constant called the polytropic index.
Quantities at the stellar center will be denoted by the subscript ``c,'' such as $ \rho_\mathrm c \equiv \rho(r=0) $ and $ P_\mathrm c \equiv P(r=0) = K\,\rho_\mathrm c^{1+\frac{1}{n}} $.
The stellar mass density and pressure can be expressed by a single profile function $ \theta $ as
\begin{equation}
\rho
= \rho_\mathrm c\,\theta^n\,,
\quad
P
= P_\mathrm c\,\theta^{n+1}\,,
\end{equation}
which is normalized as $ \theta_\mathrm c = \theta(r=0) = 1 $.
Introducing a length scale\footnote{Note that $ \ell $ depends on the polytropic index $ n $.}
\begin{equation}
\ell
\equiv
  \sqrt{\frac{(n+1)\,P_\mathrm c}{4\pi\,G\,\rho_\mathrm c^2}}\,,
\end{equation}
the non-dimensional radial coordinate $ \xi $ and graviton mass parameters $ \mu_s $ are defined as
\begin{equation}
\xi
\equiv
  \frac{r}{\ell}\,,
\quad
\mu_s
\equiv
  m_s\,\ell\,.
\end{equation}

Following the standard derivation of the Lane--Emden equation in GR, we operate the non-dimensionalized laplacian operator $ \triangle_\xi \equiv \frac{1}{\xi^2}\,\frac{\mathrm d}{\mathrm d\xi} \left(\xi^2\,\frac{\mathrm d}{\mathrm d\xi}\right) = \ell^2\,\triangle $ on the hydrostatic equilibrium condition \eqref{eq:hydro_psi} to obtain
\begin{equation}
\triangle_\xi \theta
+ \sum_s \alpha_s\,\left(
   \theta^n + \frac{\mu_s^2\,\phi_s}{4\pi G\rho_\mathrm c\ell^2}
  \right)
= 0\,.
\label{eq:hydro_theta}
\end{equation}
Unlike GR, however, this is still an integro-differential equation for $ \rho $ since $ \phi_s $ involve integrals of $ \rho $ in a non-trivial manner as given by \eqref{eq:phi}.
Since $ \phi_s $ is a formal integral of the Helmholtz equation \eqref{eq:Helmholtz}, it reduces in turn to the source term $ 4 \pi\,G\,\rho $ by an operation of the Helmholtz operator $ \triangle - m_s^2 $\,.
Thus, operating $ (\triangle_\xi - \mu_2^2)\,(\triangle_\xi - \mu_0^2) $ and using \eqref{eq:Helmholtz}, we obtain a sixth-order differential equation for $ \theta $\,:
\begin{equation}
\triangle_\xi\,\left[
 (\triangle_\xi - \mu_2^2)\,(\triangle_\xi - \mu_0^2)\,\theta
 + \alpha_2\,\left(\triangle_\xi - \mu_0^2\right)\,\theta^n
 + \alpha_0\,\left(\triangle_\xi - \mu_2^2\right)\,\theta^n
\right]
= 0\,.
\label{eq:master}
\end{equation}
This is our master equation for the profile function $ \theta $ in generic linear MG, which is an extension of the original second-order Lane--Emden equation in GR.
The stellar structure can be constructed upon integration of \eqref{eq:master} with suitable boundary conditions.
The fact that this is sixth order in differentiation is a consequence of the fact that there exist three physical degrees of freedom, one from the massive spin-$ 2 $ graviton, one from the massive spin-$ 0 $ and one from the matter.
A word of caution is that the overall laplacian operator on the left-hand side of \eqref{eq:master} cannot be dropped as in the case of field equations \eqref{eq:eom} since $ \theta $ and its derivatives must satisfy non-trivial boundary conditions at a finite radius $ r = R $, as we shall shortly see.

In terms of the non-dimensional variables, the stellar mass $ M $ and charge $ \Sigma_s $ are expressed as
\begin{equation}
M
= 4\pi\,\ell^3\,\rho_\mathrm c\,
  \int_0^{\xi_R}\!\mathrm d\xi\,\xi^2\,\theta(\xi)^n\,,
\quad
\Sigma_s
= \frac{4\pi\,\ell^3\,\rho_\mathrm c}{\mu_s}\,
  \int_0^{\xi_R}\!\mathrm d\xi\,\xi \sinh{\mu_s\xi}\,\theta(\xi)^n\,,
\label{eq:MandSigma}
\end{equation}
where $ \xi_R \equiv R/\ell $ and is the first positive zero of $ \theta $.
Unlike GR, $ M $ cannot be expressed in terms of derivatives at the stellar surface.

\subsection{Boundary conditions}

Now we move on to discuss boundary conditions to be imposed on the profile function $ \theta $.
Since the master equation \eqref{eq:master} is sixth order in differentiation, there is need for six independent conditions, a priori.
All such conditions can be derived as the requirements for the compatibility with the hydrostatic equilibrium condition \eqref{eq:hydro}.

Let us start by re-expressing \eqref{eq:hydro} in terms of $ \theta $ as
\begin{equation}
\theta'
= -\frac{\ell^{-1}}{4\pi\,G\,\rho_\mathrm c}\,
  \frac{\mathrm d\Psi}{\mathrm dr}\,,
\label{eq:hydro2}
\end{equation}
where and hereafter the prime denotes derivative with respect to $ \xi $\,.
The derivatives at the stellar center $ \theta^{(n)}_\mathrm c \equiv \frac{\mathrm d^n\theta}{\mathrm d\xi^n}(0) $ are required to be consistent with the behavior of the potential $ \Psi = \sum_s \alpha_s\,\phi_s $ evaluated there.
It can be shown that the radial acceleration at the stellar center vanishes, $ \lim_{r\to 0} -\frac{\mathrm d\Psi}{\mathrm dr} = 0 $, which implies $ \theta_\mathrm c' = 0 $ via \eqref{eq:hydro2}.
Hence, the following two boundary conditions have so far been obtained:
\begin{equation}
\theta_\mathrm c
= 1\,,
\quad
\theta'_\mathrm c
= 0\,.
\label{eq:bc_LE}
\end{equation}
These two conditions just suffice in the case of the second-order LE equation in GR.
By contrast, four more boundary conditions are required in order to solve the full sixth-order differential equation \eqref{eq:master}.
They are found from derivatives of \eqref{eq:hydro2} as
\begin{equation}
\begin{aligned}
\theta''_\mathrm c
&
= -\frac{1}{4\pi\,G\,\rho_\mathrm c}\,\lim_{r \to 0} \frac{\mathrm d^2\Psi}{\mathrm dr^2}
= \sum_s \alpha_s\,\left(-\frac{1}{3} + \frac{\mu_s^2\,\iota_s}{3}\right)\,, \\
\theta'''_\mathrm c
&
= -\frac{\ell}{4\pi\,G\,\rho_\mathrm c}\,\lim_{r \to 0} \frac{\mathrm d^3\Psi}{\mathrm dr^3}
= 0\,, \\
\theta^{(4)}_\mathrm c
&
= -\frac{\ell^2}{4\pi\,G\,\rho_\mathrm c}\,\lim_{r \to 0} \frac{\mathrm d^4\Psi}{\mathrm dr^4}
= \sum_s \alpha_s\,\left(
   -\frac{3}{5} n \theta''_\mathrm c
   - \frac{1}{5}\,\mu_s^2
   + \frac{1}{5}\,\mu_s^4\,\iota_s
  \right)\,, \\
\theta^{(5)}_\mathrm c
&
= -\frac{\ell^3}{4\pi\,G\,\rho_\mathrm c}\,\lim_{r \to 0} \frac{\mathrm d^5\Psi}{\mathrm dr^5}
= 0\,,
\end{aligned}
\label{eq:bc}
\end{equation}
where we have non-dimensionalized the constants $ I_s $ \eqref{eq:I} appearing in the potentials $ \phi_s $ such that
\begin{equation}
\iota_s
\equiv
  \frac{\mu_s\,I_s}{4\pi\,\ell^3\,\rho_\mathrm c}
= \int_0^{\xi_R}\!\mathrm d\xi\,\xi\,\theta(\xi)^n\,\mathrm e^{-\mu_s \xi}
\label{eq:iota}
\end{equation}
and iteratively applied the conditions arising from lower derivatives to those from higher ones.
It is observed that $ \theta''_\mathrm c $ and $ \theta^{(4)}_\mathrm c $ are now related to the stellar global quantities $ \iota_s $\,, which was moreover introduced to guarantee flatness at spatial infinity.
The value of $ \iota_s $ is as yet undetermined until the profile function $ \theta $ is solved.
Therefore, these expressions of the boundary values should be considered merely formal.
Namely, the problem is not formulated as a simple initial value problem as in GR, but we will have to perform some matching procedure between the boundary values at the stellar center and the integrals of the solution over the whole domain.
Note that the stellar radius $ \xi_R $ is simultaneously determined by this procedure.

It might be useful to provide an integro-differential equation equivalent to the master equation \eqref{eq:master}.
Integrating \eqref{eq:master} twice, we have
\begin{equation}
(\triangle_\xi - \mu_2^2)\,(\triangle_\xi - \mu_0^2)\,\theta
+ \alpha_2\,\left(\triangle_\xi - \mu_0^2\right)\,\theta^n
+ \alpha_0\,\left(\triangle_\xi - \mu_2^2\right)\,\theta^n
= \mu_2^2\,\mu_0^2\,\left(1 - \alpha_2\,\iota_2 - \alpha_0\,\iota_0\right)\,,
\end{equation}
where we have already utilized the boundary conditions \eqref{eq:bc}.

\subsection{The FP limit and discontinuity}

When the mass parameter $ \mu_0^2 $ goes to infinity, i.e., the non-FP parameter $ \epsilon $ goes to zero, the massive gravity theory reduces to the FP theory.
A little care has to be taken when one wants to consider such a limit because we have assumed finiteness of the graviton masses in the derivation of the full equation.
Actually, in the massive limit, the spin-$ 0 $ Yukawa potential itself must be absent from the beginning.
A secure way to go is to begin by setting $ \alpha_0 = 0 $ in \eqref{eq:hydro_theta}, which has the effect of turning off the spin-$ 0 $ Yukawa potential $ \phi_0 $\,.
Then, operating $ (\triangle_\xi - \mu_2^2) $ only, the correct master equation for the FP theory is found to be
\begin{equation}
\triangle_\xi\,\left[(\triangle_\xi - \mu_2^2)\,\theta + \frac{4}{3}\,\theta^n\right] = 0\,,
\label{eq:master_FP}
\end{equation}
where we have restored $ \alpha_2 $ to $ 4/3 $.
The reason for this being fourth order is that the helicity-$ 0 $ component of the massive spin-$ 2 $ graviton, which is absent in GR, is still at work in addition to the matter DOF.
It is interesting to note that this can also be obtained by taking a formal limit $ \mu_0^2 \to \infty $ in the master equation \eqref{eq:master} for a generic theory.

We need four boundary conditions in total, two of which are the same as the Lane--Emden condition \eqref{eq:bc_LE}.
The remaining two are obtained by setting $ \alpha_0 = 0 $ in \eqref{eq:bc} as
\begin{equation}
\begin{aligned}
\theta''_\mathrm c
&
= -\frac{1}{4\pi\,G\,\rho_\mathrm c}\,\lim_{r \to 0} \frac{\mathrm d^2\Psi}{\mathrm dr^2}
= -\frac{4}{9}(1-\mu_2^2 \iota_2)\,,\\
\theta'''_\mathrm c
&
= -\frac{\ell}{4\pi\,G\,\rho_\mathrm c}\,\lim_{r \to 0} \frac{\mathrm d^3\Psi}{\mathrm dr^3}
= 0\,.
\end{aligned}
\label{eq:bc_FP}
\end{equation}

It is useful here to consider an integro-differential equation equivalent to the fourth-order LE equation \eqref{eq:master_FP},
\begin{equation}
(\triangle_\xi - \mu_2^2)\, \theta + \frac{4}{3} \theta^n
= \frac{4}{3}\,\mu_2^2\,\iota_2
  - \mu_2^2\,.
\end{equation}
From this, one can deduce that, when the graviton mass parameter $ \mu_2 $ goes to zero, the master equation reduces to
\begin{equation}
\triangle_\xi \theta + \frac{4}{3}\,\theta^n = 0
\end{equation}
with the boundary conditions $ \theta_\mathrm c = 1\,, \theta'_\mathrm c = 0 $.
This resembles the ordinary Lane--Emden equation in GR but has a different numerical coefficient, implying that its solutions do not recover those of GR although the mass parameter $ \mu_2 $ vanishes.
This discontinuous behavior is reminiscent of the vDVZ discontinuity.

\subsection{Absence of discontinuity in non-Fierz--Pauli theories}

Last but not least, we consider the massless limit with a non-vanishing non-FP parameter $ \epsilon $.
When $ \epsilon \neq  0 $, the doubly massless limit can be taken, where both $ \mu_2 $ and $ \mu_0 $ go to zero.
Then, our master equation \eqref{eq:master} becomes
\begin{equation}
\triangle_\xi^2\,(\triangle_\xi \theta + \theta^n) = 0\,.
\end{equation}
This equation can be integrated four times and imposing the suitable boundary conditions \eqref{eq:bc} yields
\begin{equation}
\triangle_\xi \theta + \theta^n = 0\,.
\label{eq:LE}
\end{equation}
This is the same as the standard Lane--Emden equation in GR.
Thus, we can expect that the solutions to the master equation \eqref{eq:master} will recover the corresponding solutions of the standard LE equation in GR.
This proves the absence of the discontinuity in the generic, non-FP MG.

\section{\label{sec:esol}Exact solutions}

In this section, we present analytical solutions to the FP and non-FP master equations for the polytropic indices $ n = 0 $ and $ 1 $.
The stellar radius, mass, and charge are evaluated in each case.
We also discuss constraints on the graviton mass parameters.

We first recap the solutions to the LE equation \eqref{eq:LE} in GR.
The exact solutions for $ n = 0,1 $ are well known\footnote{The exact solution for $ n = 5 $ does not have a finite radius.}:
\begin{equation}
\theta_{\mathrm{LE},0}
= 1 - \frac{\xi^2}{6}\,,
\quad
\theta_{\mathrm{LE},1}
= \frac{\sin \xi}{\xi}\,.
\end{equation}
The stellar radius $ R $ and mass $ M $ in each case are
\begin{equation}
\begin{aligned}
n = 0
&
:
\quad
R_{\mathrm{LE},0}
= \sqrt 6\,\ell_0\,,
\quad
M_{\mathrm{LE},0}
= \frac{4}{3}\,\pi\,R_{\mathrm{LE},0}^3\,\rho_\mathrm c
= 8 \sqrt 6\,\pi\,\ell_0^3\,\rho_\mathrm c\,, \\
n = 1
&
:
\quad
R_{\mathrm{LE},1}
= \pi\,\ell_1\,,
\quad
M_{\mathrm{LE},1}
= \frac{4}{\pi}\,R_{\mathrm{LE},1}^3\,\rho_\mathrm c
= 4 \pi^2\,\ell_1^3\,\rho_\mathrm c\,,
\end{aligned}
\end{equation}
where we have stressed here the length scale $ \ell $ depends on $ n $.

\subsection{General structure}

All the differential equations treated in this section will be a linear homogeneous equation for $ \theta $ sharing the common form
\begin{equation}
f(\triangle_\xi)\,\theta = 0\,,
\end{equation}
where the characteristic polynomial $ f(X) $ is degree $ 2 $ in $ X $ for the FP theory or $ 3 $ for the non-FP theories.
The general solution is characterized by the set of the roots for the characteristic equation $ f(X) = 0 $.
It always has $ X_0 = 0 $ as a root, as is obvious from Eqs.~\eqref{eq:master} and \eqref{eq:master_FP}.
For each root $ X_i $\,, one finds a fundamental solution by solving $ \triangle \theta_i = X_i\,\theta_i $\,,
\begin{equation}
\theta_i
=
\left\{
\begin{aligned}
& A_i + \frac{B_i}{\xi}
& (X_i = 0) \\
& A_i\,\frac{\sinh \sqrt{X_i}\xi}{\xi} + B_i\,\frac{\cosh \sqrt{X_i}\xi}{\xi}
& (X_i > 0) \\
& A_i\,\frac{\sin \sqrt{-X_i}\xi}{\xi} + B_i\,\frac{\cos \sqrt{-X_i}\xi}{\xi}
& (X_i < 0)
\end{aligned}
\right.\,,
\end{equation}
where $ A_i $ and $ B_i $ are arbitrary real constants.
When there is no degeneracy, the general solution is just a sum of the fundamental solutions $ \theta_i $\,.
If there are degeneracies, on the other hand, special but straightforward mathematical treatments such as variation of constants will be necessary.
Later, we will take care of an example of a degenerate situation.

\subsection{Fierz--Pauli theory}

We first present the exact solutions to the Fierz--Pauli master equation \eqref{eq:master_FP} in the cases of the polytropic indices $ n = 0 $ and $ 1 $.
In this case, only the spin-$ 2 $ graviton exists.

\subsubsection{$ n = 0 $}

For $ n =  0 $, the fourth-order equation \eqref{eq:master_FP} reduces to a homogeneous linear equation
\begin{equation}
(\triangle_\xi - \mu_2^2)\,\triangle_\xi \theta = 0\,.
\end{equation}
The general solution can be found as
\begin{equation}
\theta
= A_1
  + \frac{A_2}{\xi}
  + A_3\frac{\sinh{\mu_2\xi}}{\mu_2\xi}
  + A_4\frac{\cosh{\mu_2\xi}}{\mu_2\xi}\,,
\end{equation}
where $ A_1 $\,, $ A_2 $\,, $ A_3 $\,, and $ A_4 $ are constants of integration.
The set of the boundary conditions \eqref{eq:bc_FP} determines the constants as
\begin{equation}
A_1
= 1 - A_3\,,
\quad
A_3
= -\frac{4}{3}\,\frac{(1+\mu_2\xi_R)\,\mathrm e^{-\mu_2\xi_R}}{\mu_2^2}\,,
\quad
A_2
= A_4
= 0\,.
\quad
\end{equation}
Therefore, we get the exact solution
\begin{equation}
\theta
= 1
  - \frac{4}{3}\,\frac{(1+\mu_2\xi_R)\,\mathrm e^{-\mu_2 \xi_R}}{\mu_2^2}\,
    \frac{\sinh{\mu_2\xi}-\mu_2\xi}{\mu_2\xi}\,.
\end{equation}
As anticipated, the solution is parameterized by the stellar radius $ \xi_R $\,, the value of which must be determined by numerically solving $ \theta(\xi_R) = 0 $.
The analytical expression for the gravitational potential $ \Psi $ can be found in Appendix~\ref{app:pot}.

Figure~\ref{fig:FPn0th} shows the profile function $ \theta $ for several values of $ \mu_2 $ and compares them with GR.
In the massless limit $ \mu_2 \to 0 $, the solution reduces to $ \theta = 1 - \frac{2}{9}\,\xi^2 $, which has a different coefficient from the Lane--Emden solution in GR.
As a consequence, the stellar radius shrinks from the GR value of $ \sqrt{6} $ to $ 3 / \sqrt{2} $.
On the other hand, in the massive limit $ \mu_2 \to \infty $, the solution has an infinite radius because gravity no longer works.
The profile best resembles that of GR when $ \mu_2 $ has some finite value around $ 0.4 $, but such a large value is physically unreasonable.

\begin{figure}[htbp]
\begin{center}
\includegraphics[scale=.8]{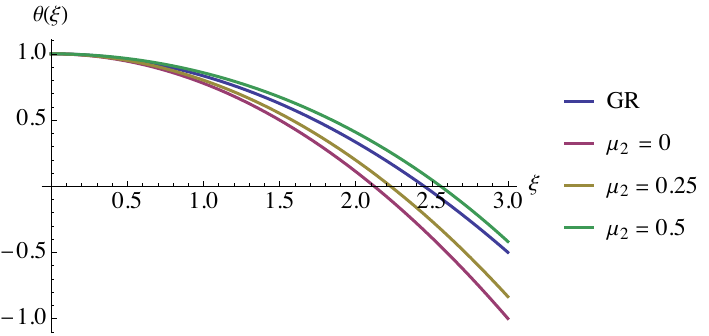}
\caption{\label{fig:FPn0th}The profile function $ \theta $ for the polytropic index $ n = 0 $ in the Fierz--Pauli theory and GR.}
\end{center}
\end{figure}

\begin{figure}[htbp]
\begin{center}
\includegraphics[scale=.8]{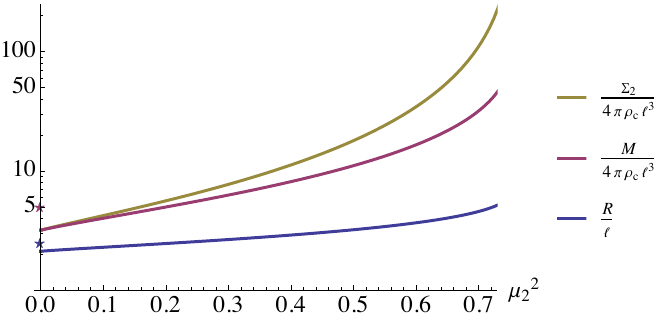}
\caption{\label{fig:FPn0vsmu}The dependence of the stellar radius $ R/\ell $ (blue), the mass $ M/(4\pi\ell^3\rho_\mathrm c) $ (red) and the charge $ \Sigma_2/(4\pi\ell^3\rho_\mathrm c) $ (yellow), each appropriately normalized, on the graviton mass $ \mu_2^2 $ for the polytropic index $ n = 0 $. The star symbols indicate the GR values of $ R $ (bottom, blue) and $ M $ (top, red).}
\end{center}
\end{figure}

The expressions for the stellar mass $ M $ and the Yukawa charge $ \Sigma_2 $ can be obtained by substituting the solution to \eqref{eq:MandSigma}:
\begin{equation}
M
= \frac{4\pi\ell^3\rho_\mathrm c\,\xi_R^3}{3}\,,
\quad
\Sigma_2
= 4\pi\ell^3\rho_\mathrm c\frac{\mu_2\xi_R\cosh{\mu_2\xi_R}-\sinh{\mu_2\xi_R}}{\mu_2^3}\,.
\label{eq:FPn0MS}
\end{equation}
Figure~\ref{fig:FPn0vsmu} shows the dependence of the stellar radius $ R/\ell $ (blue)\,, the mass $ M/(4\pi\ell^3\rho_\mathrm c) $ (red) and the charge $ \Sigma_2/(4\pi\ell^3\rho_\mathrm c) $ (yellow), each appropriately normalized, on the graviton mass parameter $ \mu_2 $\,.
As we explained earlier, when the graviton mass parameter $ \mu_2 $ goes to zero, the charge $ \Sigma_2 $ and the mass $ M $ have the same limiting value.
One also confirms that the massless limit of $ R $ and $ M $ does not converge to their corresponding GR values indicated by the star symbols.
This discontinuous behavior is analogous to what is predicted for the bending of light.
Hence, one could conclude that, in order for the graviton in the FP theory to acquire a tiny mass, some screening mechanism that helps the recovery of GR has to be invoked even inside stars.
On the other hand, as the mass of the graviton increases, each quantity diverges.
The relations between $ M $ and $ R $ and between $ \Sigma_2 $ and $ R $ can be found in Appendix~\ref{app:MR}.

\subsubsection{$ n = 1 $}

For the polytropic index $ n = 1 $, the fourth-order equation \eqref{eq:master_FP} reduces to a linear homogeneous equation
\begin{equation}
\left[\triangle_\xi-\left(\mu_2^2-\frac{4}{3}\right)\right]\triangle_\xi \theta = 0\,.
\end{equation}
We assume $ \mu_2^2 < 4/3 $ since the graviton mass should be small enough to be compatible with the gravitational-wave experiments.
The general solution is obtained as
\begin{equation}
\theta
= B_1 + \frac{B_2}{\xi}
  + B_3\frac{\sin{\lambda\xi}}{\lambda \xi}
  + B_4\frac{\cos{\lambda\xi}}{\lambda \xi}
\end{equation}
with $ \lambda = \sqrt{\frac{4}{3}-\mu_2^2} $ and $ B_1 $\,, $ B_2 $\,, $ B_3 $\,, and $ B_4 $ are constants of integration.
The boundary conditions \eqref{eq:bc_FP} determine the constants as
\begin{equation}
B_1
= 1 - B_3\,,
\quad
B_3
= \frac{4}{3\lambda^2}(1 - \mu_2^2 \iota_2)\,,
\quad
B_2
= B_4
= 0\,.
\end{equation}
Therefore, the exact solution is
\begin{equation}
\theta = 1 + B_3\,\frac{\sin{\lambda\xi} - \lambda\xi}{\xi}
\end{equation}
with
\begin{equation}
B_3
= -\frac{4(1 + \mu_2\xi_R)}
        {3\mu_2^2\,(\lambda\,\cos{\lambda\xi_R} + \mu_2\,\sin{\lambda\xi_R}) - 4\lambda\,(1 + \mu_2\xi_R)}\,.
\end{equation}
An expression for the interior gravitational potential can also be found, but we do not show it in this paper because the result is too complicated and not illuminating.

\begin{figure}[htbp]
\begin{center}
\includegraphics[scale=.8]{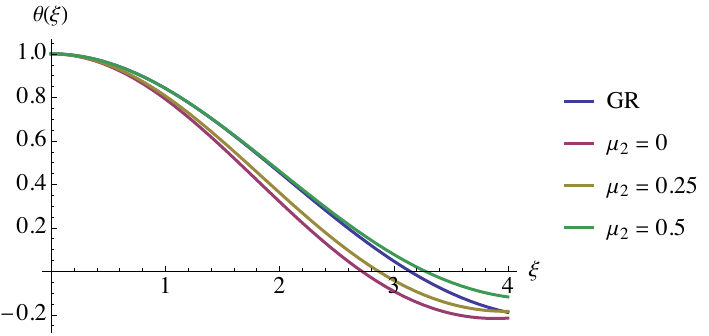}
\caption{\label{fig:FPn1th}The profile function $ \theta $ for the polytropic index $ n = 1 $ in the Fierz--Pauli theory and GR.}
\end{center}
\end{figure}

Figure~\ref{fig:FPn1th} shows the profile functions in the Fierz--Pauli theory and GR for $ n = 1 $.
Although the functional shapes are different from those for $ n = 0 $, the trends with respect to the change in the mass parameter are similar.
When $ \mu_2 \to 0 $, the solution reduces to $ \theta = \frac{\sqrt 3}{2\xi}\,\sin{\frac{2 \xi}{\sqrt 3}} $, so the stellar radius $ \xi_R $ shrinks from the GR value of $ \pi $ to $ \sqrt{3}\,\pi/2 $\,.

Once $ \xi_R $ is determined, the stellar mass $ M $ and charge $ \Sigma_2 $ can be evaluated via their expressions for $ n = 1 $:
\begin{equation}
\begin{aligned}
M
&
= \frac{4\pi\ell^3\rho_\mathrm c}{3}\,\left[
   \xi_R^3
   + \frac{3B_3}{\lambda^2}\,\left(
      \sin{\lambda\xi_R}
      - \lambda\xi_R\,\cos{\lambda\xi_R}
      - \frac{1}{3}\,\lambda^3\,\xi_R^3
     \right)
  \right]\,, \\
\Sigma_2
&
= \frac{4\pi\ell^3\rho_\mathrm c}{\mu_2^3}\,\biggl[
   \mu_2\xi_R\,\cosh{\mu_2\xi_R} - \sinh{\mu_2\xi_R} \\
& \qquad\qquad
   + B_3\,\lambda\,\left\{
      \sinh{\mu_2\xi_R}\,\left(
       1 - \frac{3}{4}\,\mu_2^2\,\cos{\lambda\xi_R}
      \right)
      - \mu_2\,\cosh{\mu_2\xi_R}\,\left(
         \xi_R + \frac{3}{4}\,\mu_2^2\,\cos{\lambda\xi_R}
        \right)\,
     \right\}
  \biggr]\,.
\end{aligned}
\label{eq:FPn1MS}
\end{equation}
The relations between $ M $ and $ R $ and between $ \Sigma_2 $ and $ R $ can be found in Appendix~\ref{app:MR}.

\subsection{Non-Fierz--Pauli generic theories}

Now we would like to tackle the full master equation \eqref{eq:master} in generic linear MG.
The existence of the two massive gravitons, spin-$ 2 $ and -$ 0 $, renders the analysis considerably messy, but most features of the solutions will be reasonably understood as collective contributions from these gravitons.
As in the FP case, we find exact solutions for the polytropic indices $ n = 0, 1 $.

\subsubsection{$ n = 0 $}

In the case of $ n = 0 $, the master equation \eqref{eq:master} reduces to
\begin{equation}
(\triangle_\xi - \mu_2^2)\,(\triangle_\xi - \mu_0^2)\,\triangle_\xi \theta = 0\,.
\end{equation}
When $ \mu_2 \neq \mu_0 $, the general solution with six arbitrary constants is found to be
\begin{equation}
\theta
= C_1 + \frac{C_2}{\xi}
  + C_3\,\frac{\sinh{\mu_2\xi}}{\mu_2\xi} + C_4\,\frac{\cosh{\mu_2\xi}}{\mu_2\xi}
  + C_5\,\frac{\sinh{\mu_0\xi}}{\mu_0\xi} + C_6\,\frac{\cosh{\mu_0\xi}}{\mu_0\xi}\,,
\end{equation}
where $ C_1 $--$ C_6 $ are constants of integration.
The boundary conditions \eqref{eq:bc} determine the constants as
\begin{equation}
C_1
= 1 - C_3 - C_5\,,
\quad
C_3
= -\frac{4}{3}(\mu_2^{-2}-\iota_2)\,,
\quad
C_5
= \frac{1}{3}(\mu_0^{-2}-\iota_0)\,,
\quad
C_2
= C_4
= C_6
= 0\,.
\end{equation}
Therefore, the exact solution is
\begin{equation}
\theta
= 1
  - \frac{4}{3}\,\frac{(1 + \mu_2\xi_R)\mathrm e^{-\mu_2\xi_R}}{\mu_2^2}\,
    \frac{\sinh{\mu_2\xi} - \mu_2\xi}{\mu_2\xi}
  + \frac{1}{3}\,\frac{(1 + \mu_0\xi_R)\mathrm e^{-\mu_0\xi_R}}{\mu_0^2}\,
    \frac{\sinh{\mu_0\xi}-\mu_0\xi}{\mu_0\xi}\,.
\label{eq:sol_n0}
\end{equation}
When $ \mu_2 = \mu_0 $, the characteristic roots degenerate and the general solution then is
\begin{equation}
\theta
= \tilde C_1 + \frac{\tilde C_2}{\xi}
  + \tilde C_3\,\frac{\sinh{\mu_2\xi}}{\mu_2\xi} + \tilde C_4\,\frac{\cosh{\mu_2\xi}}{\mu_2\xi}
  + \tilde C_5\,\sinh{\mu_2\xi} + \tilde C_6\,\cosh{\mu_2\xi}\,.
\end{equation}
A calculation leads to the solution
\begin{equation}
\theta
= 1
  - \frac{(1 + \mu_2\xi_R)\mathrm e^{-\mu_2\xi_R}}{\mu_2^2}\,
    \frac{\sinh{\mu_2\xi} - \mu_2\xi}{\mu_2\xi}\,,
\end{equation}
which happens to be identical with the formal $ \mu_0 \to \mu_2 $ limit of the non-degenerate solution \eqref{eq:sol_n0}.
The analytical expression for the gravitational potential is found in Appendix~\ref{app:pot}.

\begin{figure}[htbp]
\begin{center}
\includegraphics[scale=.8]{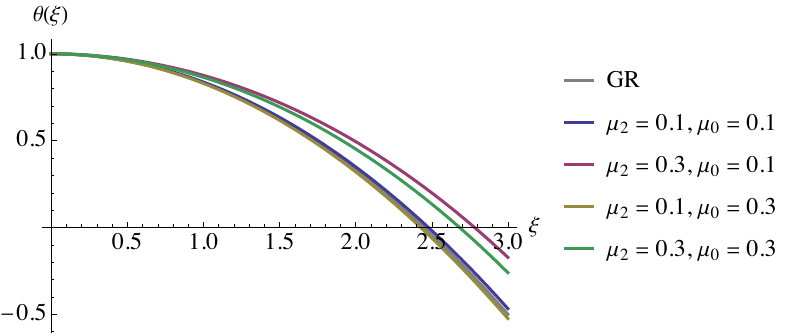}
\caption{\label{fig:n0th}Typical profile functions $ \theta $ in generic non-FP linear MG for the polytropic index $ n = 0 $ compared with the LE solution in GR.}
\end{center}
\end{figure}

Typical solutions are shown in Fig.~\ref{fig:n0th} together with the $ n = 0 $ LE solution.
As is obvious from the expression \eqref{eq:sol_n0}, the spin-$ 2 $ graviton mediates an attractive force while the spin-$ 0 $ is repulsive.
When there is a hierarchy between the graviton masses, the graviton with lighter mass dominates.
It is worth mentioning that the stellar structure is maintained even in the presence of the ghost spin-$ 0 $ graviton as long as $ \mu_0 \gtrsim \mu_2 $\,.
In the case where the graviton masses are comparable, $ \mu_2 \approx \mu_0 $\,, the spin-$ 2 $ graviton exceeds because its magnitude is four-fold stronger than the spin-$ 0 $.

Let us mention some interesting limits.
In the limit $ \mu_0 \to \infty $, i.e., $ \epsilon \to 0 $\,, the solution reduces to the one in the FP theory obtained previously.
In another special case where the spin-$ 2 $ graviton has zero mass, $ \mu_2 \to 0 $, the spin-$ 0 $ mass is simultaneously zero, $ \mu_0 \to 0 $, by definition.
The massless limit of the solution \eqref{eq:sol_n0} recovers the $ n = 0 $ LE solution $ \theta_{\mathrm{LE},0} = 1 - \xi^2/6 $, so no discontinuity appears when the spin-$ 0 $ graviton is included.

\begin{figure}[htbp]
\begin{center}
\includegraphics[scale=1]{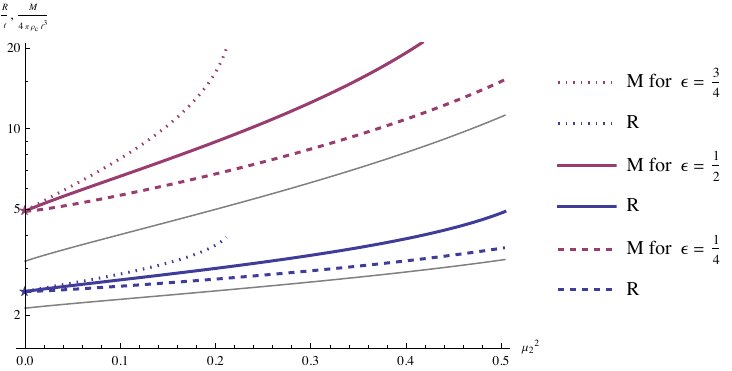}
\caption{\label{fig:plotvsmu}Dependences of the stellar radius $ R/\ell $ (blue) and the mass $ M/(4\pi\ell^3\rho_\mathrm c) $ (red), each appropriate normalized, on the graviton mass $ \mu_2^2 $ for several values of the non-FP parameter $ \epsilon $ for the polytropic index $ n = 0 $. The gray lines correspond to the FP case with $ \epsilon = 0 $. The star symbols indicate the GR values of $ R $ (bottom) and $ M $ (top).}
\end{center}
\end{figure}

Once $ \xi_R $ is determined, the stellar mass $ M $ and charge $ \Sigma_s $ can be evaluated via their expressions, which are essentially the same as the one for the FP theory \eqref{eq:FPn0MS} ($ \mu_2 \to \mu_0 $ for $ \Sigma_0 $) since $ n = 0 $.
Figure~\ref{fig:plotvsmu} shows the dependences of the stellar radius $ R/\ell $ (blue) and the mass $ M/(4\pi\ell^3\rho_\mathrm c) $ (red) on the spin-$ 2 $ graviton mass $ \mu_2^2 $ for several values of the non-FP parameter $ \epsilon $.
A larger value of $ \epsilon $ leads to a larger gradient.
Here, the Yukawa charges $ \Sigma_s $ are omitted but their behavior is similar to the stellar mass $ M $ as seen in Fig.~\ref{fig:FPn0vsmu} for the FP case.
In the presence of the spin-$ 0 $ graviton, the values of $ R $ and $ M $ in the massless limit $ \mu_2 \to 0 $ both converge to the values of GR, which is in a sharp contrast with the FP case (gray) studied in the previous subsection.

\begin{figure}[htbp]
\begin{center}
\includegraphics[scale=.8]{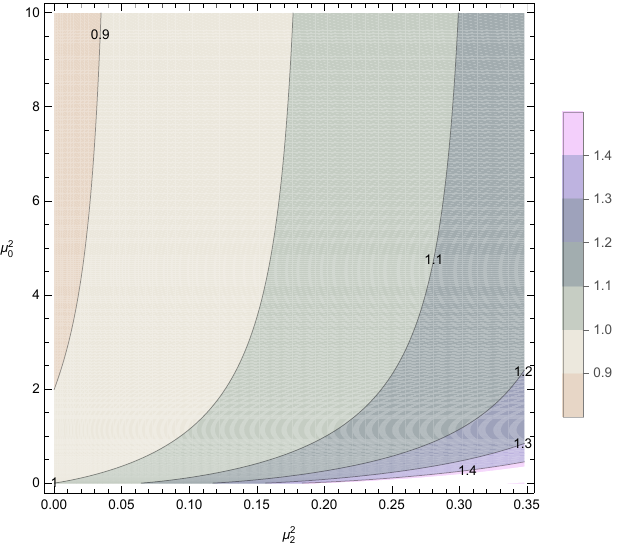}
\caption{\label{fig:lcpn0}Contours of the ratio of the stellar radius to the value of GR, $ R/R_\mathrm{LE} $\,, in the $ (\mu_2^2\,,\mu_0^2) $ plane for the polytropic index $ n = 0 $.}
\end{center}
\end{figure}

Figure~\ref{fig:lcpn0} shows contours of the ratio of the stellar radius to the value of GR, $ R/R_\mathrm{LE} $\,, in the $ (\mu_2^2\,,\mu_0^2) $ plane.
On the curve with a ratio $ R/R_\mathrm{LE} = 1 $, the MG solution has the same radius as GR as a result of competition between the attractive spin-$ 2 $ and repulsive spin-$ 0 $.
This diagram has a potential usage in constraining the graviton mass parameters.
We do not go into the detailed analysis here, but it is likely that the regions far from the line with $ R/R_\mathrm{LE} = 1 $, for instance, those with values smaller than $ 0.9 $ or greater than $ 1.1 $, may be rejected since the discrepancy from GR could be too large.
In particular, when $ \mu_2^2 \ll 1 $, the spin-$ 0 $ mass would be constrained as $ \mu_0^2 \lesssim \mathcal O(1) $.

\subsubsection{$ n = 1 $}

In the case of $ n = 1 $, the master equation \eqref{eq:master} becomes
\begin{equation}
(\triangle_\xi - X_1)\,(\triangle_\xi - X_2)\,\triangle_\xi \theta = 0\,,
\end{equation}
where $ X_i $'s are roots of the characteristic equation
\begin{equation}
f(x)
= x^2 - (\mu_2^2 + \mu_0^2 - 1)\,x + \frac{1}{3}(\mu_2^2 - 4\mu_0^2 + 3\mu_2^2\,\mu_0^2)
= 0
\end{equation}
with its discriminant being
\begin{equation}
D
= \mu_2^4 + \mu_0^4 - \frac{10}{3}\mu_2^2 + \frac{10}{3}\mu_0^2 - 2\mu_2^2\,\mu_0^2 + 1\,.
\end{equation}
In the case when $ \mu_2^2 \ll 1 $, which is of most importance on the observational grounds, the above equation has two real roots.
A further restriction $ \mu_2^2 \leq \mu_0^2 $ forces the two real roots $ X_1 $ and $ X_2 $ to have opposite signs.
The positive and negative roots then are $ \lambda_+^2 = (\mu_2^2 + \mu_0^2 - 1 + \sqrt{D})/2 $ and $ -\lambda_-^2 = (\mu_2^2 + \mu_0^2 - 1 - \sqrt{D})/2 $, respectively.
The general solution is
\begin{equation}
\theta
= D_1
  + \frac{D_2}{\xi}
  + D_3\,\frac{\sinh{\lambda_+\xi}}{\lambda_+\xi}
  + D_4\,\frac{\cosh{\lambda_+\xi}}{\lambda_+\xi}
  + D_5\,\frac{\sin{\lambda_-\xi}}{\lambda_-\xi}
  + D_6\,\frac{\cos{\lambda_-\xi}}{\lambda_-\xi}\,,
\end{equation}
where $ D_i $'s are arbitrary constants of integration.
The set of boundary conditions \eqref{eq:bc} determines the constants as
\begin{equation}
\begin{aligned}
D_1 
&
= 1 - D_3 - D_5\,,
\quad
D_2
= D_4
= D_6
= 0\,, \\
D_3 
&
= \frac{1}{\lambda_+^2\,(\lambda_+^2 +\lambda_-^2)}\,
  \left[
   1 - \lambda_-^2
   - \frac{4}{3}\,\mu_2^2\,\left(1+\iota_2\,(1-\lambda_-^2-\mu_2^2)\right)
   + \frac{1}{3}\,\mu_0^2\,\left(1+\iota_0\,(1-\lambda_-^2-\mu_0^2)\right)
  \right]\,,\\
D_5
&
= \frac{1}{\lambda_-^2\,(\lambda_+^2 +\lambda_-^2)}\,
  \left[
   1 + \lambda_+^2
   - \frac{4}{3}\,\mu_2^2\,\left(1+\iota_2\,(1+\lambda_+^2-\mu_2^2)\right)
   + \frac{1}{3}\,\mu_0^2\,\left(1+\iota_0\,(1+\lambda_+^2-\mu_0^2)\right)
  \right]\,.
\end{aligned}
\end{equation}
Therefore, the exact solution is
\begin{equation}
\theta
= 1
  + D_3\,\frac{\sinh{\lambda_+\xi} - \lambda_+\xi}{\lambda_+\xi}
  + D_5\,\frac{\sin{\lambda_-\xi} - \lambda_-\xi}{\lambda_-\xi}\,.
\end{equation}

\begin{figure}[htbp]
\begin{center}
\includegraphics[scale=.8]{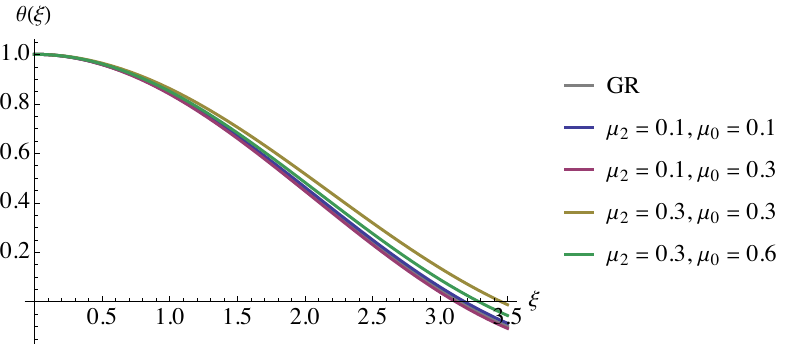}
\caption{\label{fig:n1th}The profile function $ \theta $ in non-FP generic linear MG and GR for the polytropic index $ n = 1 $.}
\end{center}
\end{figure}

Figure~\ref{fig:n1th} shows the form of the profile functions $ \theta $ for several combinations of $ \mu_2 $ and $ \mu_0 $.
The trends with respect to the changes of the values of $ \mu_2 $ and $ \mu_0 $ are almost in parallel to the case with $ n = 0 $.

\begin{figure}[htbp]
\begin{center}
\includegraphics[scale=.8]{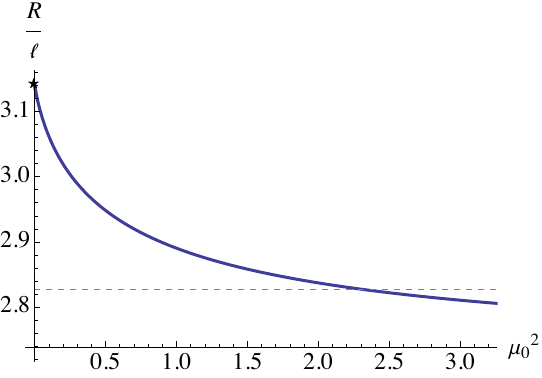}
\caption{\label{fig:n1Rvsmu0}The stellar radius $ R/\ell $ against the spin-$ 0 $ graviton mass $ \mu_0^2 $\,. The gray dashed line is the $ 90\,\% $ value of GR.}
\end{center}
\end{figure}

Here, we focus on the limiting case with the vanishing spin-$ 2 $ mass, i.e., $ \mu_2^2 \to 0 $.
Figure~\ref{fig:n1Rvsmu0} shows the dependence of the stellar radius $ R/\ell $ on the spin-$ 0 $ graviton mass $ \mu_0^2 $\,.
The first thing to note is that, as is expected, the simultaneous massless limit $ \mu_0^2 \to 0 $ recovers the GR value of $ \pi $ indicated by the star symbol in the figure.
As $ \mu_0 $ increases, the repulsive force weakens so the stellar radius shrinks.
The gray dashed line in Fig.~\ref{fig:n1Rvsmu0} indicates the $ 90\,\% $ value of the normalized radius in the case of GR, i.e., $ 0.9 \times \pi \approx 2.83 $, which we shall regard as a tentative lower bound.
If one demands the radius should be above $ 2.83 $, then $ \mu_0^2 $ should be less than about $ 2.4 $.
In this manner, at any rate, the spin-$ 0 $ mass is constrained to be $ < \mathcal O(1) $ if $ \mu_2 \ll 1 $.

\section{\label{sec:concl}Conclusions and discussion}

In this paper, we studied the structure of non-relativistic polytropic stars in the Fierz--Pauli (FP) theory and generic linear massive gravity (MG) theories.
Our aim was to study the effects of the graviton mass $ m $ and the ``non-Fierz--Pauli'' parameter $ \epsilon $ incorporated in the MG action \eqref{eq:action_g} on the stellar structures.
The spin-$ 2 $ graviton is the only content in the FP theory with $ \epsilon = 0 $, while the spin-$ 0 $ ghost graviton is also included in generic non-FP MG with $ \epsilon \neq 0 $.

In Sec.~\ref{sec:lane}, we formulated useful variables and their governing equations.
The scalar-type metric perturbations on a flat background could be neatly reorganized into the two helicity variables $ \phi_2 $ and $ \phi_0 $ as defined in \eqref{eq:phi_def}.
These variables constitute the gauge-invariant potential $ \Psi $ as in Eq.~\eqref{eq:potential} while obeying the Helmholtz-type equations \eqref{eq:Helmholtz} in static configurations.
The negative coefficient in front of $ \phi_0 $ in \eqref{eq:potential} illuminates its ghost nature.
Our first finding was the absence of the van Dam--Veltman--Zakharov (vDVZ) discontinuity for the external gravitational potential in the limit of vanishing graviton masses in non-FP MG.

Then, together with the polytropic EOS, the hydrostatic equilibrium condition \eqref{eq:hydro} lead to the modified Lane--Emden (LE) equation \eqref{eq:master} and boundary conditions \eqref{eq:bc_LE}--\eqref{eq:bc}.
The reason for the master equation being a sixth-order differential equation is the presence of the two extra gravitational DOFs.
The boundary conditions were given in terms of the derivatives at the stellar center, but involved integrals of the profile function through $ \iota_s $ \eqref{eq:iota}.
The reduced set of equations for the case of the FP theory ($ \epsilon = 0 $) was derived as in Eqs.~\eqref{eq:master_FP}--\eqref{eq:bc_FP}, where the master equation was fourth order because the spin-$ 0 $ potential $ \phi_0 $ had been turned off.
As expected, the massless limit of the master equation for the FP theory did not recover the original LE equation in GR, signaling the emergence of discontinuity analogous to vDVZ.
On the other hand, as imagined, the doubly massless limit of the non-FP equation smoothly connected to GR.

In Sec.~\ref{sec:esol}, we found exact solutions to the modified LE equations in the cases of the polytropic indices $ n = 0 $ and $ 1 $.
Typical solutions in the FP theory ($ \epsilon = 0 $) were depicted in Figs.~\ref{fig:FPn0th} ($ n = 0 $) and \ref{fig:FPn1th} ($ n = 1 $).
As these figures illustrated, the radius of a star monotonically decreases with the decreasing spin-$ 2 $ graviton mass $ \mu_2 $\,, with the minimum value reached in the massless limit being $ \sqrt 3/2 $ times the GR value.
Figure~\ref{fig:FPn0vsmu} showed the dependence of the stellar radius $ R / \ell $ (blue), mass $ M / (4\pi\ell^3\rho_\mathrm c) $ (red) and Yukawa charge $ \Sigma_2 / (4\pi\ell^3\rho_\mathrm c) $ (yellow) on $ \mu_2 $ for $ n = 0 $.
In the limit of vanishing graviton mass, $ \mu_2 \to 0 $, both $ R $ and $ M $ did not converge to the values of GR, proving the presence of discontinuities analogous to vDVZ.

Typical solutions in non-FP theories ($ \epsilon \neq 0 $) were presented in Figs.~\ref{fig:n0th} ($ n = 0 $) and \ref{fig:n1th} ($ n = 1 $).
Although the attractive spin-$ 2 $ and repulsive spin-$ 0 $ competed in these cases, we could confirm that the solutions had a finite radius within an observationally reasonable range of parameters.
Figure~\ref{fig:plotvsmu} showed the dependence of the stellar radius $ R / \ell $ (blue) and the mass $ M / (4\pi\ell^3\rho_\mathrm c) $ (red) on the spin-$ 2 $ graviton mass parameter $ \mu_2^2 $ for several values of $ \epsilon $.
The convergence to the GR values for $ \epsilon \neq 0 $ was observed, visualizing the absence of the vDVZ-like discontinuity.
The contour plot of the ratio $ R/R_\mathrm{LE} $ in Fig.~\ref{fig:lcpn0} told us that there were many possible combinations of the mass parameters that had the same radius as GR.
We argued that, if we demand that the radius must not deviate significantly from GR, the masses should be constrained to fall into a region around the $ R/R_\mathrm{LE} = 1 $ contour, say between $ 0.9 $ and $ 1.1 $.
Considering that the spin-$ 2 $ mass must be tiny, $ \mu_2 \ll 1 $, as implied by the GW experiments, the spin-$ 0 $ mass was constrained as $ \mu_0 \lesssim \mathcal O(1) $ as was read off from Figs.~\ref{fig:lcpn0} and \ref{fig:n1Rvsmu0}.

The presence of the vDVZ-like discontinuity in the massless limit of the FP theory would again imply the need for some screening mechanism so as to make the FP theory compatible with any observations which are consistent with the predictions of GR.
On the other hand, the fact that the spin-$ 0 $ ghost in generic non-FP theories can take a role in smoothly recovering GR in the doubly massless limit might suggest that the ghost offers a different mechanism to give a tiny but non-zero mass to the spin-$ 2 $ graviton.
Of course, one might think the absence of the discontinuity might be at a cost of security against possible instabilities brought by the ghost.
We leave the stability issue to future work, where our analytical solutions in this study will serve as background on which stability can be examined.

\begin{acknowledgments}
The authors are grateful to Hideki Asada for useful conversations and encouragements.
This work was in part supported by JST SPRING, Grant Number JPMJSP2152 (TT) and JSPS KAKENHI Grant Number JP23K03376 (YS).
\end{acknowledgments}

\appendix

\section{\label{sec:gauge}Gauge transformations and gauge-invariant variables}

A general metric perturbation $ h_{\mu\nu} $ about a Minkowski background can be decomposed into scalar, vector, and tensor variables as
\begin{equation}
h_{\mu\nu}\,\mathrm dx^\mu\,\mathrm dx^\nu
= -2 A\,\mathrm dt^2
  - 2\,(\partial_i B + B_i)\,\mathrm dt\,\mathrm dx^i
  + 2\,(\delta_{ij}\,C +\partial_i \partial_j E + \partial_{(i} E_{j)} + H_{ij})\,
    \mathrm dx^i\,\mathrm dx^j\,,
\end{equation}
where vector and tensor variables satisfy $ \partial_i B^i = \partial_i E^i = \partial_i H^{ij} = H_i{}^i = 0 $ and the parentheses around tensor indices denote symmetrization.
An active coordinate transformation $ x^\mu \rightarrow x^\mu + \xi^\mu(x) $ with $ \xi^\mu $ being as small as $ h_{\mu\nu} $ in magnitude transforms the metric perturbation, to first order, as
\begin{equation}
h_{\mu\nu}
\rightarrow
  h_{\mu\nu} - \pounds_\xi \eta_{\mu\nu}\,,
\end{equation}
where $ \pounds_\xi $ is the Lie derivative along $ \xi^\mu $\,.
$ \xi^\mu $ can be decomposed into the scalar and vector parts as $ (\xi^\mu) = (T,\partial^i L + L^i) $ with $ \partial_i L^i = 0 $.
It is obvious that this does not affect the tensor variable:
\begin{equation}
H_{ij}
\rightarrow
  H_{ij}\,.
\end{equation}
On the other hand, the vector variables are transformed as
\begin{equation}
B_i
\rightarrow
  B_i + \dot L_i\,,
\quad
E_i
\rightarrow
  E_i - L_i\,,
\end{equation}
where the dot denotes differentiation with respect to $ t $.
Hence, the following combination is found to be invariant:
\begin{equation}
\Sigma_i
\equiv
  B_i + \dot E_i\,.
\end{equation}
The transformations of the scalar variables are
\begin{equation}
A
\rightarrow A
  -\dot T\,,
\quad
B
\rightarrow
  B - T + \dot L\,,
\quad
C
\rightarrow
  C\,,
\quad
E
\rightarrow
  E - L\,,
\end{equation}
from which a useful set of invariant combinations is found to be
\begin{equation}
\Psi
\equiv
  A - \dot B - \ddot E\,,
\quad
\Phi
\equiv
  C\,.
\end{equation}

\section{\label{app:metric}Metric perturbations in terms of massive gravitons}

The original metric variables $ A $\,, $ B $\,, $ C $\,, and $ E $ can be expressed in terms of $ \phi_2 $ and $ \phi_0 $ as
\begin{equation}
\begin{aligned}
A
&
= \frac{4}{3m_2^4}\,\triangle^2 \phi_2
  - \left(\frac{1-\epsilon}{\epsilon}-\frac{2}{3m_2^2}\,\triangle\right)\,\phi_0
  + \frac{8 \pi\,G}{3m_2^2}\,\left(
     -3\rho
     - \frac{2}{m_2^2}\,\triangle\rho
     + \frac{2}{m_2^2}\,\triangle^2 \sigma
    \right)\,,\\
B
&
= \frac{8}{3m_2^4}\,\triangle\dot\phi_2
  + \frac{4}{3m_2^2}\,\dot\phi_0
  + \frac{16 \pi\,G}{3m_2^2}\,\left(
     3\nu
     - \frac{2}{m_2^2}\,\dot\rho
     + \frac{2}{m_2^2}\,\triangle\dot\sigma
    \right)\,,\\
C
&
= -\frac{2}{3m_2^2}\,\triangle \phi_2
  - \frac{1}{2}\,\phi_0
  + \frac{8 \pi\,G}{3m_2^2}\,(\rho-\triangle\sigma)\,,\\
E
&
= \frac{2}{m_2^2}\,\left(1-\frac{2}{3m_2^2}\triangle\right)\,\phi_2
  - \frac{2}{3m_2^2}\,\phi_0
  + \frac{8 \pi\,G}{3m_2^2}\,\left(
     3 \sigma
     - \frac{2}{m_2^2}\,\triangle\sigma
     + \frac{2}{m_2^2}\rho
    \right)\,.
\end{aligned}
\end{equation}
Therefore, the gauge-invariant variables are
\begin{equation}
\begin{aligned}
\Psi
&
= A - \dot B - \ddot E
= 2 \left(1-\frac{1}{3m_2^2}\triangle\right)\,\phi_2
  - \frac{1}{3}\,\phi_0
  + \frac{8 \pi\,G}{3m_2^2}\,(\rho-\triangle\sigma)\,,\\
\Phi
&
= C
= -\frac{2}{3m_2^2}\,\triangle \phi_2
  - \frac{1}{3}\,\phi_0
  + \frac{8 \pi\,G}{3m_2^2}\,(\rho-\triangle\sigma)\,.
\end{aligned}
\end{equation}

\section{\label{app:pot}Gravitational potentials for $ n = 0 $}

Below is a summary of the expressions for the gravitational potential for the polytropic index $ n = 0 $.
In GR,
\begin{equation}
\Psi
= \left\{
\begin{aligned}
 & -\frac{G M}{R}\,\frac{3 R^2-r^2}{2 R^2}
 & (r \leq R) \\
 & -\frac{G M}{r}
 & (r \geq R)
\end{aligned}
\right.\,, \\
\end{equation}
in the FP theory,
\begin{equation}
\Psi
= \left\{
\begin{aligned}
 & {-\frac{4}{3}}\,\frac{G\Sigma_2}{r}\,
   \frac{m_2r-(1+m_2R)\mathrm e^{-m_2R}\sinh{m_2r}}{m_2R\cosh{m_2R}-\sinh{m_2R}}
 & (r \leq R) \\
 & -\frac{4}{3}\,\frac{G\Sigma_2}{r}\,\mathrm e^{-m_2 r}
 & (r \geq R)
\end{aligned}
\right.\,, \\
\end{equation}
and in the non-FP theory,
\begin{equation}
\Psi
= \left\{
\begin{aligned}
 & {-\frac{4}{3}}\,
   \frac{G\Sigma_2}{r}\,
   \frac{m_2 r - (1 + m_2R)\,\mathrm e^{-m_2R}\,\sinh{m_2r}}{m_2R\,\cosh{m_2R}-\sinh{m_2R}}
   + \frac{1}{3}\,
     \frac{G\Sigma_0}{r}\,
     \frac{m_0 r - (1 + m_0R)\,\mathrm e^{-m_0R}\,\sinh{m_0r}}{m_0R\,\cosh{m_0R}-\sinh{m_0R}} 
 & (r \leq R) \\
 & -\frac{4}{3}\,\frac{G\Sigma_2}{r}\,\mathrm e^{-m_2r}
   + \frac{1}{3}\,\frac{G\Sigma_0}{r}\,\mathrm e^{-m_0r}
 & (r \geq R)
\end{aligned}
\right.\,.
\end{equation}

\section{\label{app:MR}Mass-to-radius and charge-to-radius relations in FP theory}

\begin{figure}[htbp]
\begin{center}
\includegraphics[scale=.7]{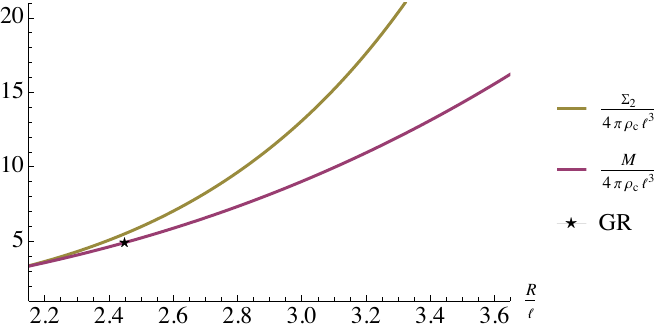}
\includegraphics[scale=.7]{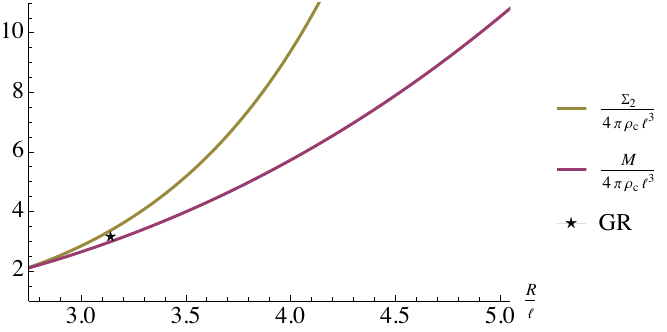}
\caption{\label{fig:FPn0MSvsR}The mass $M/(4\pi\ell^3\rho_\mathrm c)$ (red) and the charge $\Sigma_2/(4\pi\ell^3\rho_\mathrm c)$ (yellow) versus the radius $R/\ell$ for $n=0$ (left) and $1$ (right). The star symbol indicates the GR value of the mass.}
\end{center}
\end{figure}

In this Appendix, we show the relations between the mass $ M $ and radius $ R $ and between the charge $ \Sigma_2 $ and radius $ R $ in the FP theory.
The left panel of Fig.~\ref{fig:FPn0MSvsR} shows the mass $M/(4\pi\ell^3\rho_\mathrm c)$ (red) and the charge $\Sigma_2/(4\pi\ell^3\rho_\mathrm c)$ (yellow) versus the radius $R/\ell$ for the polytropic index $ n = 0 $ and the values of $ M $ and $ R $ are indicated by the star symbol.
The right panel shows the same plot but for $n=1$.
For $ n = 0 $ (left), the expression for the stellar mass $ M $ \eqref{eq:FPn0MS} is the same as GR and there is a value of $ \mu_2 $ for which both the radius and mass have the same value as in GR.
On the other hand, for $ n = 1 $ (right), the expression for $ M $ \eqref{eq:FPn1MS} is not the same as GR and the GR values cannot be realized for any $ \mu_2 $\,.

\bibliography{mgstar}

\begin{thebibliography}{10}%
\makeatletter
\providecommand \@ifxundefined [1]{%
 \@ifx{#1\undefined}
}%
\providecommand \@ifnum [1]{%
 \ifnum #1\expandafter \@firstoftwo
 \else \expandafter \@secondoftwo
 \fi
}%
\providecommand \@ifx [1]{%
 \ifx #1\expandafter \@firstoftwo
 \else \expandafter \@secondoftwo
 \fi
}%
\providecommand \natexlab [1]{#1}%
\providecommand \enquote  [1]{``#1''}%
\providecommand \bibnamefont  [1]{#1}%
\providecommand \bibfnamefont [1]{#1}%
\providecommand \citenamefont [1]{#1}%
\providecommand \href@noop [0]{\@secondoftwo}%
\providecommand \href [0]{\begingroup \@sanitize@url \@href}%
\providecommand \@href[1]{\@@startlink{#1}\@@href}%
\providecommand \@@href[1]{\endgroup#1\@@endlink}%
\providecommand \@sanitize@url [0]{\catcode `\\12\catcode `\$12\catcode
  `\&12\catcode `\#12\catcode `\^12\catcode `\_12\catcode `\%12\relax}%
\providecommand \@@startlink[1]{}%
\providecommand \@@endlink[0]{}%
\providecommand \url  [0]{\begingroup\@sanitize@url \@url }%
\providecommand \@url [1]{\endgroup\@href {#1}{\urlprefix }}%
\providecommand \urlprefix  [0]{URL }%
\providecommand \Eprint [0]{\href }%
\providecommand \doibase [0]{https://doi.org/}%
\providecommand \selectlanguage [0]{\@gobble}%
\providecommand \bibinfo  [0]{\@secondoftwo}%
\providecommand \bibfield  [0]{\@secondoftwo}%
\providecommand \translation [1]{[#1]}%
\providecommand \BibitemOpen [0]{}%
\providecommand \bibitemStop [0]{}%
\providecommand \bibitemNoStop [0]{.\EOS\space}%
\providecommand \EOS [0]{\spacefactor3000\relax}%
\providecommand \BibitemShut  [1]{\csname bibitem#1\endcsname}%
\let\auto@bib@innerbib\@empty
%</preamble>
\bibitem [{\citenamefont {de~Rham}(2014)}]{deRham:2014zqa}%
  \BibitemOpen
  \bibfield  {author} {\bibinfo {author} {\bibfnamefont {C.}~\bibnamefont
  {de~Rham}},\ }\href {https://doi.org/10.12942/lrr-2014-7} {\bibfield
  {journal} {\bibinfo  {journal} {Living Rev. Rel.}\ }\textbf {\bibinfo
  {volume} {17}},\ \bibinfo {pages} {7} (\bibinfo {year} {2014})},\ \Eprint
  {https://arxiv.org/abs/1401.4173} {arXiv:1401.4173 [hep-th]} \BibitemShut
  {NoStop}%
\bibitem [{\citenamefont {Fierz}\ and\ \citenamefont
  {Pauli}(1939)}]{Fierz:1939ix}%
  \BibitemOpen
  \bibfield  {author} {\bibinfo {author} {\bibfnamefont {M.}~\bibnamefont
  {Fierz}}\ and\ \bibinfo {author} {\bibfnamefont {W.}~\bibnamefont {Pauli}},\
  }\href {https://doi.org/10.1098/rspa.1939.0140} {\bibfield  {journal}
  {\bibinfo  {journal} {Proc. Roy. Soc. Lond. A}\ }\textbf {\bibinfo {volume}
  {173}},\ \bibinfo {pages} {211} (\bibinfo {year} {1939})}\BibitemShut
  {NoStop}%
\bibitem [{\citenamefont {van Dam}\ and\ \citenamefont
  {Veltman}(1970)}]{vanDam:1970vg}%
  \BibitemOpen
  \bibfield  {author} {\bibinfo {author} {\bibfnamefont {H.}~\bibnamefont {van
  Dam}}\ and\ \bibinfo {author} {\bibfnamefont {M.~J.~G.}\ \bibnamefont
  {Veltman}},\ }\href {https://doi.org/10.1016/0550-3213(70)90416-5} {\bibfield
   {journal} {\bibinfo  {journal} {Nucl. Phys. B}\ }\textbf {\bibinfo {volume}
  {22}},\ \bibinfo {pages} {397} (\bibinfo {year} {1970})}\BibitemShut
  {NoStop}%
\bibitem [{\citenamefont {Zakharov}(1970)}]{Zakharov:1970cc}%
  \BibitemOpen
  \bibfield  {author} {\bibinfo {author} {\bibfnamefont {V.~I.}\ \bibnamefont
  {Zakharov}},\ }\href@noop {} {\bibfield  {journal} {\bibinfo  {journal} {JETP
  Lett.}\ }\textbf {\bibinfo {volume} {12}},\ \bibinfo {pages} {312} (\bibinfo
  {year} {1970})}\BibitemShut {NoStop}%
\bibitem [{\citenamefont {Vainshtein}(1972)}]{Vainshtein:1972sx}%
  \BibitemOpen
  \bibfield  {author} {\bibinfo {author} {\bibfnamefont {A.~I.}\ \bibnamefont
  {Vainshtein}},\ }\href {https://doi.org/10.1016/0370-2693(72)90147-5}
  {\bibfield  {journal} {\bibinfo  {journal} {Phys. Lett. B}\ }\textbf
  {\bibinfo {volume} {39}},\ \bibinfo {pages} {393} (\bibinfo {year}
  {1972})}\BibitemShut {NoStop}%
\bibitem [{\citenamefont {Hinterbichler}(2012)}]{Hinterbichler:2011tt}%
  \BibitemOpen
  \bibfield  {author} {\bibinfo {author} {\bibfnamefont {K.}~\bibnamefont
  {Hinterbichler}},\ }\href {https://doi.org/10.1103/RevModPhys.84.671}
  {\bibfield  {journal} {\bibinfo  {journal} {Rev. Mod. Phys.}\ }\textbf
  {\bibinfo {volume} {84}},\ \bibinfo {pages} {671} (\bibinfo {year} {2012})},\
  \Eprint {https://arxiv.org/abs/1105.3735} {arXiv:1105.3735 [hep-th]}
  \BibitemShut {NoStop}%
\bibitem [{\citenamefont {Tachinami}\ \emph {et~al.}(2021)\citenamefont
  {Tachinami}, \citenamefont {Tonosaki},\ and\ \citenamefont
  {Sendouda}}]{Tachinami:2021jnf}%
  \BibitemOpen
  \bibfield  {author} {\bibinfo {author} {\bibfnamefont {T.}~\bibnamefont
  {Tachinami}}, \bibinfo {author} {\bibfnamefont {S.}~\bibnamefont
  {Tonosaki}},\ and\ \bibinfo {author} {\bibfnamefont {Y.}~\bibnamefont
  {Sendouda}},\ }\href {https://doi.org/10.1103/PhysRevD.103.104037} {\bibfield
   {journal} {\bibinfo  {journal} {Phys. Rev. D}\ }\textbf {\bibinfo {volume}
  {103}},\ \bibinfo {pages} {104037} (\bibinfo {year} {2021})},\ \Eprint
  {https://arxiv.org/abs/2102.05540} {arXiv:2102.05540 [gr-qc]} \BibitemShut
  {NoStop}%
\bibitem [{\citenamefont {Tonosaki}\ \emph {et~al.}(2023)\citenamefont
  {Tonosaki}, \citenamefont {Tachinami},\ and\ \citenamefont
  {Sendouda}}]{Tonosaki:2023trc}%
  \BibitemOpen
  \bibfield  {author} {\bibinfo {author} {\bibfnamefont {S.}~\bibnamefont
  {Tonosaki}}, \bibinfo {author} {\bibfnamefont {T.}~\bibnamefont
  {Tachinami}},\ and\ \bibinfo {author} {\bibfnamefont {Y.}~\bibnamefont
  {Sendouda}},\ }\href {https://doi.org/10.1103/PhysRevD.108.024037} {\bibfield
   {journal} {\bibinfo  {journal} {Phys. Rev. D}\ }\textbf {\bibinfo {volume}
  {108}},\ \bibinfo {pages} {024037} (\bibinfo {year} {2023})},\ \Eprint
  {https://arxiv.org/abs/2303.03853} {arXiv:2303.03853 [gr-qc]} \BibitemShut
  {NoStop}%
\bibitem [{\citenamefont {Abbott}\ \emph {et~al.}(2017)\citenamefont {Abbott}
  \emph {et~al.}}]{LIGOScientific:2017zic}%
  \BibitemOpen
  \bibfield  {author} {\bibinfo {author} {\bibfnamefont {B.~P.}\ \bibnamefont
  {Abbott}} \emph {et~al.} (\bibinfo {collaboration} {LIGO Scientific, Virgo,
  Fermi-GBM, INTEGRAL}),\ }\href {https://doi.org/10.3847/2041-8213/aa920c}
  {\bibfield  {journal} {\bibinfo  {journal} {Astrophys. J. Lett.}\ }\textbf
  {\bibinfo {volume} {848}},\ \bibinfo {pages} {L13} (\bibinfo {year}
  {2017})},\ \Eprint {https://arxiv.org/abs/1710.05834} {arXiv:1710.05834
  [astro-ph.HE]} \BibitemShut {NoStop}%
\bibitem [{\citenamefont {Jaccard}\ \emph {et~al.}(2013)\citenamefont
  {Jaccard}, \citenamefont {Maggiore},\ and\ \citenamefont
  {Mitsou}}]{Jaccard:2012ut}%
  \BibitemOpen
  \bibfield  {author} {\bibinfo {author} {\bibfnamefont {M.}~\bibnamefont
  {Jaccard}}, \bibinfo {author} {\bibfnamefont {M.}~\bibnamefont {Maggiore}},\
  and\ \bibinfo {author} {\bibfnamefont {E.}~\bibnamefont {Mitsou}},\ }\href
  {https://doi.org/10.1103/PhysRevD.87.044017} {\bibfield  {journal} {\bibinfo
  {journal} {Phys. Rev. D}\ }\textbf {\bibinfo {volume} {87}},\ \bibinfo
  {pages} {044017} (\bibinfo {year} {2013})},\ \Eprint
  {https://arxiv.org/abs/1211.1562} {arXiv:1211.1562 [hep-th]} \BibitemShut
  {NoStop}%
\end{thebibliography}%

\end{document}